\documentclass[12pt]{article}
\usepackage{epsfig}

\usepackage{array}

\def\appendix#1{
  \addtocounter{section}{1}
 \setcounter{equation}{0}
  \renewcommand{\thesection}{\Alph{section}}
 \section*{Appendix \thesection\protect\indent \parbox[t]{11.715cm} {#1}}
  \addcontentsline{toc}{section}{Appendix \thesection\ \ \ #1}
  }

\newcommand{\newsection}{
\setcounter{equation}{0}
\section}

\newcommand{\eq}[1]{\begin{equation} #1 \end{equation}}
\newcommand{\ar}[1]{\begin{eqnarray} #1 \end{eqnarray}}
\newcommand{\tr}{\mathop{\mathrm{tr}}\nolimits}

\def\e{{\,\rm e}\,}
\def\d{\partial}
\def\D{\delta}

\newcommand{\br}[1]{\left( #1 \right)}
\newcommand{\vev}[1]{\left\langle #1 \right\rangle}
\newcommand{\rf}[1]{(\ref{#1})}
\newcommand{\non}{\nonumber \\*}
\def\N{${\cal N}=4$ }

\def\o{{\cal O}}

\def\c{{\cal C}}
\newcommand{\bc}[1]{\left. #1 \right|_{\partial D}}

\def\cir{{\rm circle}}
\def\ep{\varepsilon}

\begin{document}
\begin{titlepage}
\begin{flushright}
ITEP--TH--63/01
\end{flushright}
\vspace{.5cm}

\begin{center}
{\LARGE Wilson loops in SYM theory:} 
\\[.25cm]
{\LARGE from weak to strong coupling}\\
\vspace{.9cm} {\large Gordon
W. Semenoff$\,^1$\footnote{semenoff@nbi.dk ~~Permanent address:
Department of Physics and Astronomy, University of British Columbia,
6224 Agricultural Road, Vancouver, B.C. V6T 1Z1 Canada.  Work
supported by NSERC of Canada and SNF of Denmark.}  and
K. Zarembo$\,^2$\footnote{Konstantin.Zarembo@teorfys.uu.se ~~Also
at ITEP, B. Cheremushkinskaja, 25, 117259 Moscow, Russia.  Work
supported by the Royal Swedish Academy of Sciences and by STINT grant
IG 2001-062. }}\\
\vspace{24pt}
{\it $^1$ The Niels Bohr Institute}
\\{\it Blegdamsvej 17, DK2100 Copenhagen \O, Denmark}
\\ \vskip .2 cm
{\it $^2$ Department of Theoretical Physics}\\
{\it Uppsala University, Box 803, S-751 08 Uppsala, Sweden}
\end{center}
\vskip .25cm
\begin{abstract} 
\noindent
We review Wilson loops in ${\cal N}=4$ supersymmetric Yang-Mills
theory with emphasis on the exact results.
The implications are discussed in the context of
the AdS/CFT correspondence.
\end{abstract}

\noindent
This is a review compiled from presentations given by
the authors at:
{\it Sapporo Winter School}, Sapporo, Japan, January,
2002;
{\it APCTP/KIAS Winter School}, Seoul/Pohang, Korea, December
2001;
{\it 14th Nordic Network  Meeting}, Stockholm, November 2001;
{\it ``Light-cone Physics: Particles and Strings''}, Trento, 
Italy, September, 2001;
{\it ``Particles, Fields and Strings''}, Burnaby, British Columbia, July 2001;
{\it Tohwa Symposium}, Fukuoka, Japan, July, 2001;
{\it APS DPF Northwest Meeting}, Seattle, May 2001;
{\it MRST Meeting}, London, Ontario, May, 2001;
{\it Symposium on Gauge Theory}, Jena, Germany,
February, 2001; {\it Lake Louise Winter School}, February, 2001;
{\it Canadian Institute for Advanced Research Meeting},
Banff, Alberta, February, 2001.

\end{titlepage}


\setcounter{page}{2}

\newsection{The AdS/CFT correspondence}

One of the most remarkable aspects of string theory is the existence
of a dual description of D-branes.  In perturbative string theory
D-branes are D+1-dimensional hypersurfaces in spacetime where open
strings are allowed to begin and end.  On the other hand, they are
also identified with the solitonic black brane solutions of
supergravity or type II superstring theory.  This gives two
alternative formulations of their dynamics.  In the first, the low
energy degrees of freedom are described by gauge fields which are the
lowest energy excitations of the open strings.  The dynamics is that
of a supersymmetric Yang-Mills gauge theory living on the world-volume
of the branes.  In the second, the low energy dynamics is supergravity
which is the low energy limit of closed string theory. The degrees of
freedom are fluctuations of the supergravity fields about the black
brane background and they live in the bulk of ten dimensional
spacetime.

There are some situations where these two descriptions have an
overlapping domain of validity.  In those cases, the same physical
system is described by two different theories which must therefore be
dual to each other.  Because the degrees of freedom in these theories
live on spaces of different dimensions, this has been called
holographic duality, and is often viewed as an explicit realization of
old ideas about the degrees of freedom in quantum gravity
~\cite{'tHooft:gx}\cite{Susskind:1994vu}.  The application of
holographic duality to study the relationship between gauge fields and
gravity is known as the AdS/CFT correspondence
~\cite{Petersen:1999zh}-\cite{D'Hoker:2002aw}.

The most precise statement of holographic duality is contained in the
Maldacena conjecture \cite{Maldacena:1998re}.  In this conjecture, on
the gravity side, the asymptotically flat exterior of an extremal
black D3-brane is replaced by its near-horizon geometry which is a
product of 5-dimensional anti-de Sitter (AdS) space and the 5-sphere,
$AdS^5\times S^5$.  The conjecture in its strongest form then asserts
an exact duality between type IIB superstring theory on this
background and four dimensional ${\cal N}=4$ supersymmetric Yang-Mills
theory (SYM) on flat 4-dimensional space.  The gauge group of the
Yang-Mills theory is $SU(N)$ and there are $N$ units of Ramond-Ramond
(RR) 4-form flux in the string theory.  This duality includes a
prescription for identifying correlation functions in the two theories
~\cite{Gubser:1998bc}\cite{Witten:1998qj}.

${\cal N}=4$ supersymmetric Yang-Mills theory has vanishing beta
function and is a conformal field theory.  Its degrees of freedom are
a gauge field $A_\mu$, six scalars $\Phi_i$ and four Majorana spinors
$\Psi$.  All fields transform in the adjoint representation of the
gauge group.  The Lagrangian (in Euclidean space) is
\begin{eqnarray}
{\cal L}=\frac{1}{g_{YM}^2}{\rm Tr}
\left\{ \frac{1}{2}F_{\mu\nu}^2+(D_\mu\Phi_i)^2-\sum_{i<j}[\Phi_i,\Phi_j]^2+
i\bar\Psi\Gamma^\mu D_\mu\Psi+i \bar\Psi\Gamma^i[\Phi_i,\Psi]\right\}.
\end{eqnarray}
This action can be obtained as a dimensional reduction of
ten-dimensional ${\cal N}=1$ supersymmetric Yang-Mills theory to four
dimensions.  This is reflected in our notation where we assemble the
fermions into a single ten-dimensional 16-component Majorana-Weyl
spinor $\Psi$ with $(\Gamma^\mu,\Gamma^i)$ the ten dimensional Dirac
matrices in the Majorana-Weyl representation.

The dual supergravity background is the near-horizon geometry of a
black D3-brane which has $N$ units of RR-flux. This is the string
theory state with  $N$ coinciding D3-branes. The metric can be
written with coordinates $(x^\mu,y^i)$, $\mu=1,...,4$, $i=1,...,6$ in
the form
\begin{equation}
ds^2= R^2 \, \frac{dx^\mu dx^\mu+ dy^i dy^i}{y^2}.
\label{metric}
\end{equation}
The unit 6-vector $y^i/y$ parameterizes $S^5$ and $x^\mu,y$ are the
coordinates of $AdS^5$.  The $AdS^5$ and $S^5$ have equal radii of
curvature, $R$.  The boundary of the space is at $y=0$ and the AdS
horizon is at $y=\infty$.  The metric written explicitly in product
form is
\begin{equation}
ds^2=R^2\frac{dx_\mu^2+dy^2}{y^2}+R^2d\Omega_{S^5}^2.  
\end{equation}
In the AdS/CFT correspondence, 
the radius $R$ is related to the Yang-Mills coupling by
\begin{equation}
R=\sqrt{\alpha'} \left( g_{YM}^2N\right)^{1/4}
\end{equation}
 The line-element (\ref{metric}) is invariant under coordinate
transformations which form the AdS group $SO(2,4)$.  The
rest of the isometry group of (\ref{metric}) is the symmetry group of
$S^5$, which is $SO(6)\sim SU(4)$. Together with the supersymmetry, these
form the super-group $SU(2,2|4)$.

On the gauge theory side, the bosonic symmetries are manifest as the
$SO(2,4)$ conformal symmetry and the $SU(4)$ R-symmetry of SYM theory.
In fact, the $SO(2,4)$ transformations which leave the AdS metric
(\ref{metric}) invariant reduce to conformal transformations on the
boundary of $AdS^5$ where the SYM observables are defined.  The radial
coordinate $y$ is associated with the scale in the SYM theory
\cite{Akhmedov:1998vf,Peet:1998wn} -- larger objects on the boundary
probe larger distances in $AdS^5$.

The string theory on the background metric (\ref{metric}) is a sigma
model with coupling constant given by the inverse of the effective
string tension,
\begin{equation}
T=R^2/2\pi\alpha'=\sqrt{g_{YM}^2N}/2\pi
\end{equation} 
which is a dimensionless quantity.

Furthermore, the string coupling $g_s$ and the Yang-Mills coupling
$g_{\rm YM}$ are related by 
\begin{equation}
g_s=4\pi g_{\rm YM}^2
\end{equation}
This relation can be understood from the fact that the gauge theory
action, in front of which the gauge coupling should appear as the
factor $1/g_{\rm YM}^2$, is obtained from the disc amplitude in string
perturbation theory which is of order $1/g_s$.

With these identifications, the string theory and the SYM theory are
conjectured to be exactly equivalent. This equivalence is a remarkable
and extremely non-trivial fact.  However, it is hard to work out its
consequences in the general setting, when both sides of the duality
correspond to complicated strongly interacting systems.  Weaker and
computationally more useful versions of the AdS/CFT duality are
obtained by taking limits (table~1).  The 't~Hooft limit of the gauge
theory \cite{'tHooft:1974jz} takes $g_{YM}\to0$ and
$N\rightarrow\infty$ with the 't~Hooft coupling $\lambda\equiv
g_{YM}^2N$ held fixed.  In the string theory, this coincides with the
classical limit where $g_s\rightarrow 0$ and the radius of curvature
of the background space, $R$, is held constant.  In this limit, large
$N$ Yang-Mills theory is dual to classical string theory on
the $AdS^5\times S^5$ background.

\begin{table}[h]
\caption{\small Different limits of the AdS/CFT correspondence}
\label{dregimes}
\begin{center}
\begin{tabular}{|c|c|}
\hline
{\Large \bf $\bf N=4$ SYM} & {\Large\bf String theory in 
$\bf AdS^5\times S^5$}\\
\hline
Yang-Mills coupling: $g_{YM}$ & String coupling: $g_s$ \\
Number of colors: $N$ & String tension: $T$ \\
\hline
\multicolumn{2}{|c|}{\bf Level 1: Exact equivalence} \\
\hline
\multicolumn{2}{|c|}{$g_s=g^2_{YM}/4\pi,~~~~~T=\sqrt{g^2_{YM}N}/2\pi$} \\
\hline
\multicolumn{2}{|c|}{\bf Level 2: Equivalence in the 't Hooft limit} \\
\hline
$N\rightarrow\infty,~~~\lambda=g^2_{YM}N$-fixed & $g_s\rightarrow 0,
~~~T$-fixed \\
(planar limit) & (non-interacting strings) \\
\hline
\multicolumn{2}{|c|}{\bf Level 3: Equivalence at strong coupling} \\
\hline
$N\rightarrow\infty,~~~\lambda\gg 1$ & $g_s\rightarrow 0,~~~T\gg 1$ \\
& (classical supergravity) \\
\hline
\end{tabular}
\end{center}
\end{table}

The $g_s\to 0$ limit of string theory in AdS space is still a
complicated dynamical theory.  The limit projects the string path
integral onto an integration over world-sheets of minimal genus.  In
this limit, the string sigma model is still a highly non-linear two
dimensional conformal field theory.  This sigma model simplifies in
its weak coupling limit, which coincides with the limit where the
string tension $T$ and hence the radius of curvature of the space in
string units is taken to be large.  When the string tension is large,
only massless states of the string are important.  Other states become
infinitely massive and decouple from low energy physics. Thus, the
limit of the type IIB string theory which takes the string tension
to infinity is approximated by type IIB supergravity
on the background $AdS^5\times S^5$.  In the gauge theory, this
corresponds to the limit of large 't~Hooft coupling
$\lambda\to\infty$.  Thus, the strongly coupled large $N$ limit of
Yang-Mills theory should coincide with IIB supergravity on the
background $AdS^5\times S^5$.
  
Even the last, weakest version of this duality has profound
consequences.  Previous to it, the main quantitative tool which could
be used for super-Yang-Mills theory was perturbation theory in
$g_{YM}^2$, the Yang-Mills coupling constant.  This is limited to the
regime where $g_{YM}^2$ and $\lambda$ are both small.  The conjectured
duality allows one to do concrete computations in a new regime, the
limit where $g_{YM}^2$ is small and $N$ and $\lambda$ are both large
\cite{Gubser:1998bc,Witten:1998qj}.

The large $N$ expansion of gauge theory has long been thought to be
related to some sort of weakly coupled string theory
\cite{'tHooft:1974jz}.  Development of this idea has been limited by
the fact that, although some qualitative features of the large $N$
limit are known, it is not possible to solve the infinite $N$ limit
explicitly.  Maldacena's conjecture now gives one explicit example
where a string theory is dual to a gauge theory.
Moreover,  the string theory
can be used to solve the large $N$ and large $\lambda$ limit of the
gauge theory.

The best evidence in support the AdS/CFT correspondence comes from
symmetry.  The global symmetries on the both sides of the correspondence
combine into the super-group
$SU(2,2\mid 4)$.  Not
only are the global symmetries the same, but some of those objects
which carry the representations of the symmetry group --- the spectrum
of chiral operators in the field theory and the fields in supergravity
theory can be matched \cite{Witten:1998qj}.  Furthermore, both
theories are conjectured to have a Montonen-Olive $SL(2,Z)$ duality.

The super-conformal symmetry of ${\cal N}$=4 super-Yang-Mills theory
severely restricts the form of correlation functions and in some cases
it protects them from radiative corrections so that they have only a
trivial dependence on the coupling constant.  A number of these have
been computed using the AdS/CFT correspondence and have been found to
agree with their free field limit. This can be viewed as a
simultaneous confirmation of supersymmetric non-renormalization
theorems and the prediction of AdS/CFT.  Examples are the two- and
three-point functions of chiral primary operators ~\cite{Lee:1998bx}.
 
However, because AdS/CFT and perturbation theory compute different limits, it
is difficult to obtain an explicit check of the AdS/CFT correspondence
for a quantity which has a non-trivial dependence on the coupling
constant.  An example of such a quantity is the free energy of
Yang-Mills theory heated to temperature $T$ which, because of
conformal invariance, must be of the form
\begin{equation}
F= -f(\lambda,N)\frac{\pi^2}{6}N^2T^4V
\label{freeenergy}
\end{equation}
When computed perturbatively in the large $N$ limit,
$$f(\lambda,N)=1-3\lambda/2\pi^2+\ldots$$ The gravitational dual of SYM at 
non-zero temperature is the
AdS black hole with Hawking temperature $T$.  Its free energy can be
deduced from its Beckenstein-Hawking entropy.  There are also stringy
corrections computed in \cite{Gubser:1998nz,Pawelczyk:1998pb}.
The result is (\ref{freeenergy}) with
$$f(\lambda,N)=\frac{3}{4}+
\frac{45}{32}\zeta(3)\lambda^{-\frac{3}{2}}+\ldots$$ The first
computation is an expansion in $\lambda$ whereas the second is an
expansion in $1/\lambda^{1/2}$.  Though it is not known in the
intermediate regime, it has been conjectured that $f(\lambda)$ is a
smooth function that interpolates monotonically between 1 and
$\frac{3}{4}$ as $\lambda$ goes from 0 to $\infty$.  The corrections
on both sides go in the right direction and are consistent with
monotonicity of the transition from weak to strong coupling.

There are now a few examples of quantities which are non-trivial
functions of the coupling constant and whose large $N$ limit is
computable and is thought to be known to all orders in perturbation
theory in planar diagrams
\cite{Erickson:2000af,Drukker:2000rr,Semenoff:2001xp}.  All of these
quantities involve Wilson loops, which play an important role in the
AdS/CFT correspondence for several reasons.  Apart from allowing one to
obtain exact results in certain cases, Wilson loops are the 
objects in \N SYM theory whose string theory dual is a source for strings.
Thus, they  probe string theory directly.  This is true even in the
supergravity regime where the string that is induced by a Wilson loop source 
behaves as a classical object.
A review of Wilson loops in \N SYM theory is the central theme of this
Paper.  This review is not comprehensive.  Our main emphasis will be
on the exact results and we will omit several interesting issues which
are discussed extensively elsewhere.  Notable omissions are
computation of quantum corrections to Wilson loops due string
fluctuations \cite{Dru00',For99,Gre98,Nai99,Kin00,Forste:2001ah}, the
instanton contribution to Wilson loop expectation values
\cite{Bianchi:2001jg,Bianchi:2002gz}, and extensions to less
supersymmetric and non-conformal examples of gauge theory / gravity
correspondence. Wilson loops in the that context are reviewed in
\cite{Sonnenschein:1999if}.

\newsection{Wilson loops at strong coupling}

The Wilson loop operator in ${\cal N}=4$ super-Yang-Mills theory is
associated with the holonomy of a heavy W-Boson.  This
W-Boson arises when the $SU(N+1)$ gauge symmetry is broken to
$SU(N)\times U(1)$ and the symmetry breaking condensate
is  sent to infinity. The phase factor in the path-integral representation
of the W-Boson propagator involves not only gauge fields,  but also
scalars:
\eq{\label{php} W(C)=\frac{1}{N}\,\tr{\rm P}\exp\left[\oint_C
d\tau\,\br{i A_\mu(x)\dot{x}^\mu +\Phi_i(x)\theta^i|\dot{x}|}\right].
} Here, $C$ is a closed curve parameterized by $x^\mu(\tau)$ and
$\theta_i$ is a unit 6-vector, $\theta^2=1$, in the direction of the
symmetry breaking condensate. 

This operator plays more important role in the AdS/CFT
correspondence than the usual Wilson loop 
 for several reasons. One of  the most important of them
is supersymmetry.
 The supersymmetry transformations of
gauge and scalar fields are
\begin{equation}
\delta_\epsilon A_\mu(x)=\bar\Psi\Gamma_\mu\epsilon,~~~~~
\delta_\epsilon \Phi_i(x)=\bar\Psi(x)\Gamma_i\epsilon
\end{equation}
Under the infinitesimal supersymmetry transformation, 
the exponent in the Wilson loop changes by
 $$\bar\Psi\left(i\Gamma_\mu\dot x^\mu(\tau)-\Gamma_i\theta^i|\dot
x(\tau)|\right)\epsilon.$$
The linear combination of Dirac matrices
$\left(i\Gamma_\mu\dot x^\mu(\tau)-\Gamma_i\theta^i|\dot
x(\tau)|\right)$ squares to zero and has eight zero
eigenfunctions. When these
eigenfunctions are $\tau$-independent, 
the loop retains half of the supersymmetry.
This occurs
only when $\dot x^\mu(\tau)$ is a constant, that is, when $C$ is a
straight line.  In that case 
 $W(C)$ is a BPS
operator  that commutes with half of the supercharges.
Consistent with this property, it seems to be protected from radiative
corrections. 
Indeed, in the leading orders of perturbation theory 
and also in the strong coupling limit which is
computed by the AdS/CFT correspondence, it is independent of the
coupling constant and  
\begin{equation}
\vev{W({\rm straight~line})}=1
\label{straightline}
\end{equation}
 A Wilson loop which
is not a straight line but is a smooth curve still has local
supersymmetry and has better ultraviolet properties than the
conventional loop which does not have the scalar field.

The AdS/CFT correspondence can be used to compute the expectation
value of a Wilson loop in the large $\lambda$, large $N$ limit.
In Yang-Mills theory, the amplitude for a heavy W-boson 
to traverse a
closed curve $C$ of length $L(C)$ is given by the vacuum expectation
value of the Wilson loop
accompanied by an 
exponential factor which is associated with the mass of the W-Boson:
\begin{equation}
{\cal A} = \e^{-ML(C)}\vev{W(C)},
\label{amplitude}
\end{equation}
where $M$ is the mass and this formula is accurate when $M\to\infty$.  

According to the AdS/CFT correspondence, this amplitude can also be computed
using string theory.  The strings propagate in the bulk of $AdS^5\times S^5$ 
and we should consider those whose worldsheets have boundary on the
loop $C$ \cite{Maldacena:1998im,Rey:1998ik}:
\ar{\label{equality}
{\cal A}&=&\int DX^\mu DY^i Dh_{ab} D\vartheta^{\alpha}
\exp
\left(-\frac{\sqrt{\lambda}}{4\pi}\,\int_D
d^2\sigma\,\sqrt{h}\,h^{ab}\,\frac{
\partial_a X^\mu\partial_b X^\mu+\partial_a Y^i\partial_b Y^i}{Y^2}
\right.
\non
&&\left.
\vphantom{
-\frac{\sqrt{\lambda}}{4\pi}\,\int
d^2\sigma\,\sqrt{h}\,h^{ab}\,\frac{
\partial_a X^\mu\partial_b X^\mu+\partial_a Y^i\partial_b Y^i}{Y^2}
} 
+{\rm fermions}\right),
\label{strpartfn}
}
where $\vartheta^\alpha$ are anticommuting
coordinates on the superspace whose bosonic part is 
$AdS^5\times S^5$.
The fermion piece of the world sheet action, which makes it supersymmetric, 
is known \cite{Metsaev:1998it,Kallosh:1998zx,Pesando:1998fv}
and takes a reasonably simple form 
in a suitable gauge \cite{Kallosh:1998nx,Kallosh:1998ji}, but
we will not need its explicit form here.
The contour $C$ is located on the boundary of $AdS^5$, 
and the string partition function is supplemented by the following
boundary conditions: 
\eq{\bc{X^\mu}=x^\mu(\tau),~~\bc{Y^i}=\theta^i\bc{Y},~~\bc{Y}=0.  \label{bc}}

The string partition function (\ref{strpartfn}) defines a complicated
2 dimensional sigma model which cannot be solved exactly. It
simplifies considerably in the large 't~Hooft coupling limit where the
string tension, $T=\sqrt{\lambda}/2\pi$, becomes large and suppresses
string fluctuations.  The superstring path integral is then dominated
by the bosonic action at its saddle-point. The saddle-point
corresponds to a minimal surface in $AdS^5\times S^5$.  Because of the
$O(6)$ symmetry of the boundary condition (\ref{bc}), the minimal
surface is embedded in $AdS^5$ and sits at a particular point,
$\theta^i$ on $S^5$.

The string action at the saddle-point is obtained by minimizing
the Nam\-bu-Go\-to action, that is classically equivalent to
the Polyakov action in \rf{equality}:
$$
{\rm Area}(C)= \int d^2\sigma\,\frac{1}{Y^2}\sqrt{\det_{ab}
\br{\d_a X^\mu\d_b X^\mu+\d_a Y\d_b Y}}.
$$
If we equate the two vacuum amplitudes (\ref{amplitude}) and (\ref{equality})
and solve for the Wilson loop we get
\begin{equation}
-\ln \vev{W(C)}= \frac{\sqrt{\lambda}}{2\pi}{\rm Area}(C) -ML(C).
\label{unreg}
\end{equation}

The area of a surface whose boundary is $C$ is infinite.  This
infinite part should cancel between the terms on the right-hand-side
of (\ref{unreg}) when we take $M$ to infinity. We shall discuss the
reason for this cancellation shortly.  The infinite part of the area
can be regularized by letting the curve $C=\left(
x^\mu(\tau),y^i(\tau)\right)$ lie in the bulk of $AdS^5\times S^5$ and
later projecting it onto the boundary by taking $y^i(\tau)\to 0$. 
Let us show that the divergence that arises in this limit is always proportional to the 
perimeter of $C$. Take, for simplicity, $y^i(\tau)=\theta^i\ep$. Then, it is 
straightforward to solve for the minimal surface near the boundary.
In appropriate coordinates:
\eq{
Y^i(\tau,y)=y\theta^i,~~~~~
X^\mu(\tau,y)=x^\mu(\tau)+{\cal O}(y^2),
}
so
\ar{
{\rm Area}(C)&=& \int d\tau\int_{\ep} dy\, \frac{1}{y^2}
\sqrt{\dot{X}^2+\dot{X}^2X'^2-(\dot{X}\cdot X')^2}
\non
&=&\int d\tau\int_{\ep}\frac{dy}{y^2}\br{\sqrt{\dot{x}^2}
+O(y^2)}=\frac{1}{\ep}\,L(C)+{\rm finite}.
\label{areaunreg}
} 
This is the divergent part of the area which should cancel the term
with the mass of the W-Boson in (\ref{unreg}).  Since the divergent
piece is inversely proportional to the distance from the boundary,
when we take the minimal area to be a functional of the boundary
curve, the divergent part can be idenditied using the operator
$$- \oint_C d\tau\,
y^{i}(\tau)\frac{\delta}{\delta y^i(\tau)}. 
$$  
The finite part of the area determines the Wilson loop expectation value:
\begin{equation}
-\ln \vev{W(C)}=\frac{\sqrt{\lambda}}{2\pi}\, \lim_{|y|\to 0}\left( 1 +\oint_C
d\tau\,y^{i}(\tau)\frac{\delta}{\delta y^i(\tau)}\right)
{\rm Area}(C)\equiv\frac{\sqrt{\lambda}}{2\pi}\,
\hat A(C)
\label{ahat}
\end{equation}
This is a Legendre transform with respect to the variable $y^i/y^2$
which was noticed and given an interpretation in terms of T-duality in
\cite{Drukker:1999zq}. If one defines the momentum variable
$$\pi^i=-y^2\,\frac{\delta {\rm Area}[x^\mu,y^i]}{\delta y^i(\tau)},$$
 then the above
equation states that $$-\ln \vev{W(C)}= \frac{\sqrt{\lambda}}{2\pi}\,
\hat A[x^i,\pi^i]$$ is a function of
the coordinates $x^i$ and momenta $\pi^i$.
The latter should be specified
in such a way that the position of world sheet boundary, which is obtained
from it by the functional derivative
$$\frac{y^i}{y^2}=-\frac{\delta \hat A}{\delta\pi^i(\tau)}$$ is at the
boundary of the AdS space.  Of course, the equations of motion for the
variational problem with area $\hat A(C)$ are identical to those for
${\rm Area}(C)$ and the boundary condition is usually easily
implemented once $\hat A(C)$ is identified.

Let us also clarify why is it legitimate, at least on the qualitative level, 
to identify $1/\ep$ with the mass of the W-Boson.
In type II string theory, ${\cal N}=4$ super-Yang-Mills
theory describes the low energy limit of $N$ parallel D3-branes
stacked on top of each other. A
W-boson appears in the Higgs phase when the $SU(N+1)$ symmetry is
broken to $SU(N)\times U(1)$ by a condensate $\vev{\Phi_i}$.  This
corresponds to the state where one of the D3-branes is separated from
the remaining stack.  The W-boson is the lowest energy excitation of
the superstring which connects the separated brane and the stack.
Its mass is given by the string's minimal length divided by $\alpha'$.
In the full D3-brane solution of type IIB supergravity, the near-horizon
geometry, which is $AdS^5\times S^5$, is glued to the asymptotically flat region
at the boundary of AdS space.
The infinite mass of the W-boson is
proportional to the distance from the horizon to the boundary.  
The subtraction in (\ref{unreg}) is a regulated version
of this distance times the length of the contour, $C$. Indeed, the area
of the surface
\eq{\label{ads2}
Y^i=y\theta^i,~~~~~X^\mu=x^\mu(\tau),
}  
where $y$ runs from $\ep$ to infinity is exactly
$1/\ep$. The divergence in \rf{areaunreg}
 is then equal to the mass of the W-boson times $L(C)$. 
Thus, there is
exact cancellation of the subtracted term and the W-boson mass.  

By definition, the minimal surface has the smallest area for given boundary
conditions. The area of the surface (\ref{ads2}), to be subtracted for the sake
of regularization, is always larger. 
Consequently, the renormalized area is
always negative.  Thus, there are three universal predictions of the
AdS/CFT correspondence: in the strong 't~Hooft coupling limit the Wilson
loop expectation value exponentiates, the exponent is proportional to
$\sqrt{\lambda}$ and the co-efficient is positive,
\begin{equation}\label{generalstring}
\vev{W(C)}=\exp
\left(\sqrt{\lambda}\times{\rm positive~number}\right).
\end{equation}  

Corrections to the Wilson loop in the large $\lambda$ limit
come from the string fluctuations
and are suppressed when $\lambda$ is large. An expansion
which includes them perturbatively is an ordinary $\alpha'$ expansion
of the world-sheet sigma model and, for AdS string, goes in 
powers of $1/\sqrt{\lambda}$.
There is also an overall factor associated with zero modes that arises
upon gauge fixing in the integral over internal metrics.
The number of zero modes is equal to three times the Euler character
of the world sheet \cite{Alvarez:1982zi,Dru00'} and the path integral
over each zero mode contributes a factor of $\lambda^{1/4}$. For the
disk amplitude, which determines the Wilson loop expectation value in
the $g_s\to0$ limit, since the Euler character of the disk is $-1$,
this gives a factor of $\lambda^{-3/4}$. Hence, a general form of the
strong-coupling expansion for a Wilson loop expectation value is
\eq{\label{adswl1}
\vev{W(C)}=\lambda^{-3/4}\e^{-\frac{\sqrt{\lambda}}{2\pi}\hat
A(C)}\sum_{n=0}^{\infty} c_n\lambda^{-n/2}, } where $c_n$ are
numerical coefficients that depend on the contour $C$.

There are several cases of curve $C$ for which the minimal area can be
calculated explicitly.  As a warm-up exercise, we could try to produce
the conjectured expectation value of the straight-line Wilson loop
(\ref{straightline}).  There,
$$
x^\mu(\tau)=(\tau,0,0,0)
$$
By symmetry, we expect that the minimal surface which has this boundary is 
an infinite plane which is perpendicular to the boundary of AdS space,
\begin{equation}
X^\mu(\sigma,\tau)=(\tau,0,0,0)
~~,~~
Y^i(\sigma,\tau)=\sigma \theta^i
\label{minstrln}
\end{equation}
Indeed, it is easy to see that this surface solves the equations for a minimal
surface which are obtained from the area using a variational principle.  The
induced metric is
$$
ds^2= \frac{ d\tau d\tau + d\sigma d\sigma }{ \sigma^2}
$$
which is that of the space $AdS^2$.
The area element is 
$$
dA= \frac{1}{\sigma^2}d\sigma d\tau
$$
which, to compute the area should be integrated over
$\tau\in(-\infty,\infty)$ and $\sigma\in[0,\infty)$.  The integration
has two sources of divergence, one coming from the infinite length of
the line, which is $L=\int d\tau$, and the other coming from the
expected singular behavior of the area element near the boundary of
AdS space which we cut off according to our prescription of replacing
0 in the lower limit of the integral over $\sigma$ with $\epsilon$.
Then, the area is
$$
A({\rm straight~line})= \frac{L}{\epsilon}
$$
The subtracted area is
\begin{equation}
\hat A({\rm straight~line})=\left(1+\epsilon\frac{\partial}{\partial\epsilon}
\right)A({\rm straight~line})=0
\label{minstrln1}
\end{equation}
which vanishes.  Exponentiation gives (\ref{straightline}), which is
the expected unit expectation value of the straight line Wilson
loop.

Another important example is the rectangular Wilson loop.  
The interaction potential for a $W-\bar W$ static pair of
$W$-boson sources can be extracted from the expectation value of a
rectangular Wilson loop with length $T$ and width $L$ by taking the
limit \eq{
V(L)=-\lim_{T\rightarrow\infty}\frac{1}{T}\,\ln\vev{W(C_{L\times T})}.
}  Because of scale invariance, the expectation value of a rectangular
loop can depend only on the ratio $T/L$.  Then, dimensional analysis
implies that $V(L)\sim 1/L$ which is the scale invariant Coulomb
interaction.  This is what is expected to occur in a conformally
invariant gauge theory.
Indeed, solving for the minimal surface and evaluating its area one finds
\cite{Maldacena:1998im,Rey:1998ik}: \eq{
V(L)=-\frac{4\pi^2\sqrt{\lambda}}{\Gamma^4(1/4)\,L}.  } The effective
Coulomb charge turns out to be proportional to $\sqrt{\lambda}$, which
is smaller than one would expect from the naive extrapolation of the
weak-coupling $O(\lambda)$ behavior.  This can be interpreted as a
screening effect of the processes corresponding to the sum of all
planar Feynman diagrams.

\subsection{Circular loop}

Another example where the minimal area can be easily found is that of
a circular loop.

The minimal surface whose boundary at $y=0$ is a circle of radius $a$ is very simple
\cite{Berenstein:1999ij,Drukker:1999zq}.  It is the solution of the quadratic equation
\eq{\label{cir1}
x^2_1+x^2_2+y^2=a^2.
}
The induced metric of this minimal surface is  
$$
ds^2=\frac{a^2}{y^2(a^2-y^2)}\,dy^2+\frac{a^2-y^2}{y^2}\,d\varphi^2,
$$
where we parameterize the surface
by the AdS scale $y$ and the polar angle 
in the $(x_1,x_2)$ plain $\varphi$.
The area element is:
$$dA=\frac{a}{y^2}\, dy d\varphi ,$$
and the regularized minimal area
is readily computed:
\eq{\label{cirar}
\hat{A}(\cir)=\left( 1+\ep\frac{\partial}{\partial\ep}\right)
2\pi a\int_\ep^a\frac{dy}{y^2}=-2\pi.
}
For the expectation value of the circular loop we get:
\eq{\label{cir2}
W(\cir)=\e^{\sqrt{\lambda}}.
}

This result does not look suspicious, unless one wonders how it was
originally derived.  The easiest way to solve for the minimal surface
is to use the conformal invariance \cite{Berenstein:1999ij}: the
inversion transformation $x_\mu\rightarrow x_\mu/x^2$ maps the circle
onto a straight line, for which the minimal surface in
(\ref{minstrln}) is really simple.  This transformation is conformal
and can be extended to an isometry of AdS space: 
\ar{\label{inv}
x_\mu&\rightarrow&\frac{x_\mu}{x^2+y^2}\,, \non
y&\rightarrow&\frac{y}{x^2+y^2}\,.  } 
The minimal surface bounded by a straight line is a half-plane which
extends to the horizon and has a geometry of AdS$^2$.  The combination
of the inversion and the translation by $a$ in $x_1$ maps the
half-plane $x_3=0,~x_1=1/2a$ onto the hemisphere \rf{cir1}.

The minimal area for the straight line (\ref{minstrln1}), after
the divergence is removed, is zero. 
This differs from the result for the circle (\ref{cirar}).
What is
surprising is that the expectation values for the circle and for the
straight line are not the same, in apparent contradiction with the
conformal symmetry.  Since the expectation values are different for
conformally equivalent operators, conformal invariance has been violated.

The violation of conformal symmetry obviously stems from the
necessity of regularizing the area. Any regularization explicitly
breaks conformal invariance.  There is the question of whether
conformal symmetry is restored once the infinity is subtracted and the
regularization is removed.  When properly defined, the area is
finite, but renormalization amounts to subtraction of a linearly divergent
constant and this leaves a finite effect that breaks conformal invariance.
In this respect, the difference between the Wilson line and the
circular Wilson loop is reminiscent of the usual conformal anomaly.

\begin{figure}[h]
\hspace*{4cm}
\begin{center}
\epsfxsize=9cm
\epsfbox{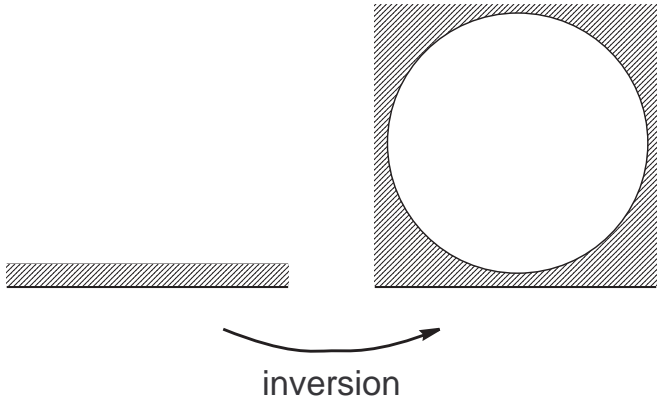}
\end{center}
\caption[x]{\small Before the conformal transformation, the regularization
cuts the slice of thickness $\ep$ near the boundary. After the transformation,
regularization cuts the exterior of the sphere of radius $1/\ep$.}
\label{regads}
\end{figure}

The simplest regularization, used in \rf{cirar}, moves the boundary
 of AdS space from $y=0$ to $y=\ep$.  The
transformation \rf{inv} maps the true boundary $y=0$ on itself and acts on
it as the inversion. But the shifted boundary $y=\ep$ gets mapped
onto a sphere of a very large radius: \eq{
x^2_\mu+\left(y-\frac{1}{2\ep}\right)^2=\frac{1}{4\ep^2}.  } Therefore,
the conformal transformation changes the regularization prescription,
fig.~\ref{regads}.
The "regularized" AdS space is now the interior of this sphere.  If we want
to calculate the minimal area for the circle by first mapping it 
onto a straight line, we must use this unusual regularization.  The
regularized minimal surface is then the interior of a circle
$$x^2+\left(y-\frac{1}{2\ep}\right)^2=\frac{1}{4\ep^2}-\frac{1}{4a^2}$$
in AdS$_2$. The discrepancy between the circle and the straight line
derives from the difference in regularization prescriptions.
 This difference becomes even more evident after
the rescaling $(x,y)\rightarrow (x/2\ep,y/2\ep)$, which is an isometry
of AdS$_2$.  The radius of the circle then becomes finite: \eq{
x^2+(y-1)^2=1-\ep^2/a^2.  } The area of its interior is \eq{
\int\frac{dx dy}{y^2}=2\int_{-\sqrt{1-\ep^2/a^2}}^{\sqrt{1-\ep^2/a^2}}
dx\,\frac{\sqrt{1-\ep^2/a^2-x^2}}{x^2+\ep^2/a^2}=\frac{2\pi
a}{\ep}-2\pi, } in agreement with \rf{cirar}.

The difference between the Wilson loop expectation values has the
classic form of an anomaly.  In both cases there is a linear
divergence that must be subtracted according to some prescription.
The subtraction is what ruins the formal symmetry which would
otherwise relate them. However, the area anomaly affects only extended
objects such as Wilson loops and should not be confused with ordinary
conformal anomaly which affects local operators and which is absent in
\N SYM in flat Euclidean space.

We have already noted that it could be expected that the straight line
Wilson loop has cancelling radiative corrections due to the fact that
it is a BPS operator, i.e. it commutes with half of the supercharges.
In the super-conformal algebra, besides the 16 supercharges, there are
also 16 superconformal charges.  The circular Wilson loop commutes
with half of the super-conformal charges.  In order to regulate the
theory, it is necessary to introduce an ultraviolet cutoff.  It is
possible to cut off in a way that does not break the supersymmetry
\cite{Erickson:2000af} and thereby preserve the algebra of
supercharges in the cut off theory.  However, the algebra of conformal
supercharges cannot be preserved since the conformal invariance is
broken by a cut off.  Thus, one might expect that the straight line
Wilson loop is more protected by supersymmetry than the circular
Wilson loop.  This leaves open the possibility that the circular loop
can get quantum corrections.

\subsection{Operator product expansion (OPE)}

When probed from a distance much larger than the size of
the loop, the Wilson loop 
should behave effectively as
a local operator.  
More precisely, it
can be expanded in a series of
local operators 
\cite{Shifman1980:ui,Berenstein:1999ij}:
\begin{equation} 
W(C)= \langle W(C)\rangle\sum
{\cal C}_A R^{\Delta_A} {\cal O}^A(0) 
\label{expansion}
\end{equation} 
where ${\cal O}^A(0)$ is an operator evaluated at the center of
the loop, $\Delta_A$ is the conformal dimension of ${\cal O}^A(x)$,
$R$ is the radius of the loop, and $\c_A$ are OPE coefficients.

The OPE coefficients can be read off from the correlation functions of the Wilson loop
with local operators. We can choose the basis of unit normalized
primary operators (those which have the lowest dimension in a given representation
of the conformal group):
\begin{equation}
\langle {\cal O}^A(x) {\cal O}^B(y) \rangle = \frac{ \delta^{AB} }{
\left| x-y \right|^{\Delta_A + \Delta_B} }
\label{normalization}
\end{equation}
Their OPE coefficients can be extracted from the large distance
behavior of the connected two-point correlator:
\eq{\label{ope}
\frac{\vev{W(C)\,\o^A(L)}_c}{\vev{W(C)}}=\c_A\,
\frac{R^{\Delta_A}}{L^{2\Delta_A}}+\ldots\,
} 
where $L\gg R$. The omitted terms correspond to descendants
and are of higher order in
$R/L$.

\begin{figure}[h]
\hspace*{4cm}
\begin{center}
\epsfxsize=12cm
\epsfbox{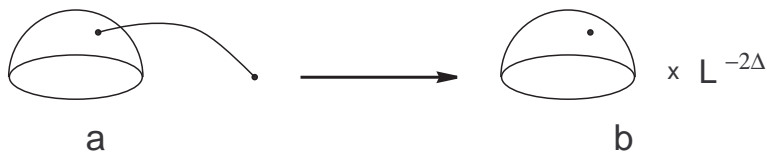}
\end{center}
\caption[x]{\small (a) A correlation function of the Wilson loop with
a local operator is determined by an exchange of the supergravity mode
between the classical string world sheet and the point of operator
insertion on the boundary of $AdS^5$.  (b) At large distances the
correlator factorizes, and the OPE coefficient is given by an integral
of the appropriate vertex operator over the world sheet.}
\label{opefig}
\end{figure}

The strong-coupling evaluation of the OPE coefficients
\cite{Berenstein:1999ij} involves a hybrid of the string and the
supergravity calculations: The classical string world sheet created by
the Wilson loop absorbs the supergravity mode emitted at the point of
operator insertion: \eq{ \frac{\vev{W(C)\,\o^A(L)}_c}{\vev{W(C)}}=\int
d^2\sigma\, \sqrt{h}\, V_A(X,\d/\d X)G_A(X,L), } where $V_A(X,\d/\d
X)$ is the vertex operator of the supergravity mode associated with
$\o^A$, $G_A(X,L)$ is bulk-to-boundary propagator, and the integral
runs over the classical string world sheet. The propagator factorizes
at large separation and gives a factor $1/L^{2\Delta_A}$
(fig.~\ref{opefig}). Indeed, the scalar bulk-to-boundary propagator
associated with
a dimension-$\Delta$ operator behaves at large distances  as\footnote{An
overall coefficient is chosen to unit normalize the two-point function.}:
\eq{
G(x,y;L)=\sqrt{\frac{\Delta-1}{2\pi^2}}\left[\frac{y}{y^2+(L-x)^2}\right]^\Delta
\rightarrow\sqrt{\frac{\Delta-1}{2\pi^2}}\,\frac{y^\Delta}{L^{2\Delta}}\,.
}
The OPE coefficient of a scalar operator 
is thus given by an integral of
 the vertex operator over the string world sheet:
\eq{
\c_A=R^{-\Delta_A}\sqrt{\frac{\Delta_A-1}{2\pi^2}}\int d^2\sigma\,
\sqrt{h}\,V_A(X,\d/\d X)Y{\Delta_A} .
}
Explicit calculations for a number of chiral operators
can be found in Ref. \cite{Berenstein:1999ij}. 

A chiral primary operator (CPO) is a primary operator which commutes
with half of the supercharges and therefore lies in a short
representation of the super-conformal algebra.  This particularly
interesting set of operators are traces of the scalar fields,
\begin{equation}\label{cpo}
{\cal O}^I_k=\frac{(8\pi^2)^{k/2}}{\sqrt{k}\lambda^{k/2}}\,
C^I_{i_1\ldots i_k}\tr \Phi^{i_1}\ldots\Phi^{i_k},
\end{equation}
where $C^I_{i_1\ldots i_k}$ are totally symmetric traceless
tensors which are normalized as
\begin{equation}
C^I_{i_1\ldots i_k}C^J_{i_1\ldots i_k}=\delta^{IJ}.  
\end{equation}
 Here, we are
following the convention of
refs.~\cite{Lee:1998bx,Berenstein:1999ij}.
The first of the CPOs,
\begin{equation}\label{dim2}
{\cal O}^{ij}=\frac{8\pi}{\sqrt{2}\,\lambda }\,\tr \left(\Phi^i\Phi^j
-\frac{1}{6}\,\delta^{ij}\Phi^2\right),
\end{equation}
has lowest possible conformal dimension, $\Delta=2$,  and in
this sense is the most important operator in \N SYM theory.

The overall coefficient in the definition of CPOs has been chosen
to unit normalize their two-point functions.  The two-point
correlators of CPOs are protected by supersymmetry and do not receive
radiative corrections.  This insures that they have the correct
normalization to all orders of perturbation theory once the
normalization is set at weak coupling. This will be important when we
will compare perturbative calculations 
to the
supergravity predictions 
for strong coupling behavior.

The AdS duals of CPOs are particular linear combinations of spin-zero
Kaluza-Klein modes on $S^5$ of the metric and the anti-symmetric two
form.  
Each CPO thus is associated with a
spherical function:
\begin{equation}
Y^I(\theta)=C^I_{i_1\ldots i_k}\theta^{i_1}\ldots\theta^{i_k}.  
\end{equation}
The OPE coefficients of a Wilson loop must be proportional to
 $Y^I(\theta)$.  An explicit calculation for the circular contour
 gives the large $\lambda$ limit of the correlator,
 \cite{Berenstein:1999ij}:
\begin{equation}\label{sc}
\frac{\langle W(C)\,{\cal O}^I_k\rangle }{\langle W(C)\rangle}
=2^{k/2-1}\sqrt{k\lambda}\,\,\frac{R^k}{L^{2k}}\,Y^I(\theta)
~~~(\lambda\rightarrow \infty).
\end{equation}

\subsection{Wilson loop correlator}\label{wlc}

The two-point correlator of Wilson loops in the regime when the distance
between the loops is large compared to their sizes
is one of the cases in which the use of OPE expansion
is justified.
For identical loops of opposite orientation separated
by distance $L$,
\begin{equation}\label{2lope}
\frac{  \vev{ W(C_1)W(C_2)}_c}{
\vev{W(C_1)}\vev{W(C_2)} }=
\sum
|{\cal C}_A|^2\left( \frac{R}{L}\right)^{2\Delta_A}\,.
\label{conn}
\end{equation}
This representation of the Wilson loop correlator imposes 
certain constraints on the
OPE coefficients. Since
 the number of operators of a given conformal dimension
 grows exponentially with the increase of the dimension,
the sum over all operators in
intermediate states in (\ref{conn}) will diverge at 
distances comparable to the size of the
loops $R\sim L$, unless operators of large dimensions are strongly
suppressed (stronger than exponentially). If there is no suppression,
the correlator of Wilson loops will undergo a phase transition at some
$L\propto R$.

Suppression of operators with large quantum numbers (such as conformal
dimensions, spins, {\it etc.})  is quite a general statement, which
applies to confining theories as well \cite{Zarembo:1999bu}. Indeed,
consider the spectral representation for the Wilson loop correlator:
\eq{ \vev{ W(C_1)W(C_2)}_c=\int_0^\infty
dE\,\rho_C(E)\e^{-EL}, } where \eq{ \rho_C(E)=\sum_{n\neq
0}\D(E-E_n)\left|\langle 0| W(C) | n\rangle\right|^2.  } 
The density
of states
is expected to grow exponentially as $\exp(E/T_H)$, where $T_H$ is the Hagedorn
temperature. The form-factor of the Wilson loop must suppress
this growth. Otherwise,
the correlator will undergo a phase transition at
$L=1/T_H$, which is similar to the Hagedorn transition
 at finite temperature. Such  phase transitions in  
correlation functions are expected in quantum gravity 
\cite{Aharony:1998tt},
but not in gauge theories. Consequently, the form-factor of the Wilson loop
must suppress highly excited states, either in \N
SYM or in confining gauge theories.

The operators of large scaling dimension are indeed suppressed at weak
coupling. Consider perturbative calculation of the OPE coefficients as
defined by eq.~\rf{ope}; say, for chiral primary operators
\rf{cpo}. The lowest order diagrams for the dimension-$k$ operator
contain at least $k$ scalar propagators that go from the operator
insertion to the Wilson loop and require an expansion of
the loop to at least $k$-th order in the scalar fields. Since
Wilson loop is an exponential, the OPE coefficient will be suppressed
by $1/k!$. The same is obviously true at weak coupling
for any other operator that has
large scaling dimension or spin.

Can we see this suppression on the supergravity side of the AdS/CFT
correspondence?  The answer to this question 
seems to be negative.  The OPE
coefficients for the dimension-$k$ chiral primary actually grow with $k$ 
at strong coupling! This follows from the  AdS/CFT prediction, eq.~ \rf{sc}.
 Does this mean that, if the coupling is
strong enough, the pair correlator of Wilson loops indeed diverges at
short distances and there is a phase transition at some critical
separation between the loops? We will argue later that growth of OPE
coefficients with dimension is an artifact of taking the
strong-coupling limit. Exact  
OPE coefficients rapidly
decrease with $k$ if we carefully take the limit $k\rightarrow\infty$
at any large but finite $\lambda$.

\begin{figure}[h]
\hspace*{4cm}
\begin{center}
\epsfxsize=10cm
\epsfbox{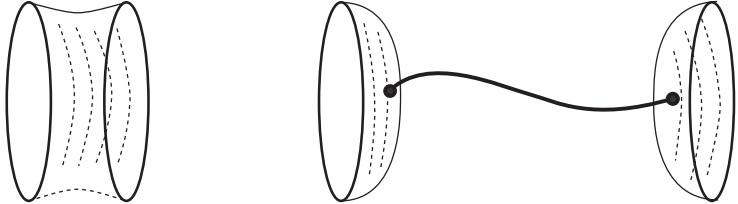}
\end{center}
\caption[x]{\small String breaking.}
\label{stbr}
\end{figure}

\begin{figure}[h]
\hspace*{4cm}
\begin{center}
\epsfxsize=8cm
\epsfbox{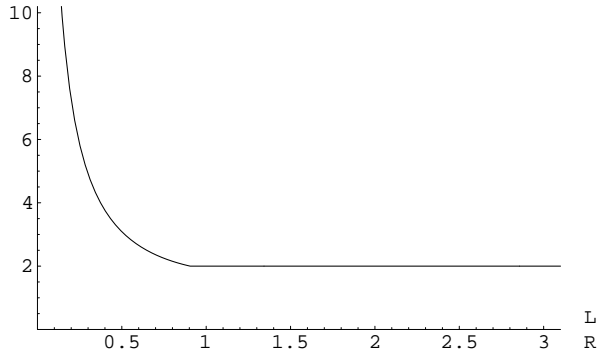}
\end{center}
\caption[x]{\small $\ln\protect\vev{W(C_1)W(C_2)}$ vs. the distance between 
the loops $C_1$ and $C_2$ for concentric circles of radius $R$ \cite{Olesen:2000ji}.
The Gross-Ooguri phase transition is of the first order and takes place at 
$L_c=0.91 R$ \cite{Zarembo:1999bu}.}
\label{GO}
\end{figure}

But there is indeed a phase transition in the Wilson loop correlator at
strong coupling. However, it is associated with another phenomenon,
the string breaking. The string breaking is a consequence of the area
law, and is specific to string theory.  At short distances, the
correlator of two Wilson loops is saturated by the string stretched
between the contours.  When the separation between 
the loops grows, the area of the string world sheet evidently
grows too.  Since the string has
tension, eventually the world sheet breaks into two minimal surfaces
that span each of the contours separately (fig.~\ref{stbr})
\cite{Gross:1998gk}.  In between the surfaces, the string world sheet
degenerates into an infinitely thin tube which describes propagation
of individual supergravity modes.  The OPE expansion \rf{2lope} then
becomes a good approximation.  The two regimes are separated by the
Gross-Ooguri phase transition, and the correlator is not analytic in
the distance between the loops
\cite{Zarembo:1999bu,Olesen:2000ji,Kim:2001td}.  As an example, we
plot the logarithm of the correlator of two circular loops as a
function of the distance $L$ between them in fig.~\ref{GO}.  The first
derivative of the correlator is discontinuous at the critical separation,
so Gross-Ooguri transition is first order in this case.

The Gross-Ooguri transition in an inherently stringy phenomenon and
looks rather counterintuitive from the field theory perspective.
Indeed, any Feynman diagram that contributes to the Wilson loop
correlator is an analytic function of the distance between the loops.
Of course, one has to sum an infinite series of all planar diagrams
to reach the strong-coupling limit on the field theory side. Surprisingly,
even partial resummation that takes into account only planar graphs
without internal vertices reveals the Gross-Ooguri transition at
strong coupling \cite{Zarembo:2001jp}. 
It is also possible to see how the Gross-Ooguri transition
disappears as one gradually decrease the coupling
on the string side of the correspondence \cite{Zarembo:1999bu}.
The string
fluctuations, that should be taken into account beyond
the strong-coupling limit,
make the transition smooth, it becomes
a crossover at finite $\lambda$
 and is completely washed 
out at weak coupling.

%

\newsection{Wilson loops in perturbation theory}

To the leading order in perturbation theory,
\begin{equation}
\langle W(C)\rangle=1+\frac{\lambda}{16\pi^2}\oint_C
d\tau_1\,d\tau_2\frac{\vert\dot x(\tau_1)\vert \vert \dot
x(\tau_2)\vert-{\dot x}(\tau_1)\cdot \dot x(\tau_2) }{ \vert
x(\tau_1)-x(\tau_2)\vert^2} +\cdots.
\label{leadloop}
\end{equation}
The first term in the integral comes from the scalars and the second
comes from vector exchange.
For a loop without cusps or self-intersections, their sum is finite.
This cancellation occurs because of local supersymmetry of the Wilson 
loop operator. Cusps and
self-intersections of the contour lead to divergences as discussed in  \cite{Drukker:1999zq}.
An expectation value for a smooth contour is known to be finite at two \cite{Erickson:2000af}
and three \cite{Plefka:2001bu} loops. The
the cancellations  are likely to persist to higher orders of perturbation theory,
though no rigorous proof of the finiteness  
to all orders has been given. 

The integrand in \rf{leadloop} is non-negative by triangle inequality.
The extremal case is the straight line, for which the correction is strictly zero,
as it should be for a BPS operator. For any other contour, 
\eq{
\ln\vev{W(C)}=\lambda\times{\rm positive~number}.
}
 This is a general prediction of perturbation theory.
Comparing to the string theory prediction \rf{generalstring}, we see that
a Wilson loop expectation value
interpolates between linear and square-root scaling with
$\lambda$ as we go from weak
to strong coupling. One would expect that this interpolation is smooth.
In particular,
 higher-order perturbative corrections should decrease $\ln\vev{W(C)}$. 
Explicit calculations indeed demonstrate that next-to-leading order corrections go
in the right direction for simplest contours. For instance, the first 
perturbative correction to the
static potential is repulsive:
\eq{
V(L)=-\br{\frac{\lambda}{4\pi}-\frac{\lambda^2}{8\pi^3}\,\ln\frac{1}{\lambda}+\ldots} 
\frac{1}{L}\,.
}
The non-analytic dependence on $\lambda$ is a consequence of an IR
divergence in the rectangular Wilson loop in the limit when its temporal
extent becomes infinite \cite{Appelquist:es}.
Careful treatment of this divergence requires infinite resummation
of Feynman diagrams, which
removes the IR singularity, but makes the static potential non-perturbative
beyond the leading order of weak coupling expansion
\cite{Erickson:2000qv,Erickson:2000af}.

Another example  is the circular loop, for which
\eq{\label{nlo}
\ln \vev{W(\cir)}=\frac{\lambda}{8}-\frac{\lambda^2}{384}+\ldots\,.
}
Again perturbative series is sign-alternating.
Diagram calculations that lead to this formula can be generalized
to include a particular class of diagrams to all orders
of perturbation theory, namely diagrams without 
internal vertices (rainbow graphs). 
The sum of these diagrams is believed to give a large-$N$
exact result for the circular Wilson loop.

\newsection{Exact results for circular Wilson loop}

As we discussed before, the circular Wilson loop is almost a 
BPS operator. The circular loop and the straight line, which is exactly BPS,
are conformally equivalent. This equivalence is spoiled by an anomaly and 
the circular loop gains an expectation value, which is a non-trivial function
of the 't~Hooft coupling. Still, one can anticipate that supersymmetry
leads to many cancellations among quantum correction for the circle.
It was argued \cite{Drukker:2000rr} that rainbow diagrams
exhaust all correction that survive supersymmetry cancellations.

\subsection{Expectation value to all orders in perturbation theory}

In this section, we consider a circular Wilson loop, whose radius we
can assume to be unity. A convenient parameterization of this loop is
$
x(\tau)=(\cos \tau, \sin \tau, 0, 0)
$.

We will sum all planar diagrams which have no
internal vertices. It is instructive to consider first the lowest order 
of perturbation theory \rf{leadloop}. 
For the circular loop, that expression greatly simplifies,
because
$$\vert
x(\tau_1)-x(\tau_2)\vert^2=2 
-2{x}(\tau_1)\cdot x(\tau_2)=2\left(1-{\dot x}(\tau_1)\cdot \dot x(\tau_2)\right),
$$
and, consequently,
\eq{\label{contr}
\frac{\vert\dot x(\tau_1)\vert \vert \dot
x(\tau_2)\vert-{\dot x}(\tau_1)\cdot \dot x(\tau_2) }{ \vert
x(\tau_1)-x(\tau_2)\vert^2}=\frac{1}{2},
}
independently of $\tau_1$ and $\tau_2$. The contour integrals 
in \rf{leadloop} are trivial and just give an
overall factor of $(2\pi)^2$. Computation of the first term in perturbative series
\rf{nlo} turns out very simple. The only complication we encounter at higher
orders is path ordering and necessity to keep only planar diagrams. 
In virtue of \rf{contr}, the gluon and the scalar propagators, whose ends lie on the 
same circle, 
always combine to  a constant. This observation makes
the problem of resummation of rainbow diagrams essentially zero-dimensional.
In fact,
we can express the circular loop in terms of a correlator in a zero-dimensional
field theory:
\eq{
\vev{W(\cir)}=\vev{\frac{1}{N}\,\tr\e^M}_M,
}
where the "path integral" is defined by the partition function
\eq{\label{rmm}
Z=\int d^{N^2}M\, \exp\br{-\frac{8\pi^2}{\lambda}\,N{\rm tr} M^2}.
}
Averaging over $M$ correctly accounts for the 
combinatorics of rainbow diagrams and the measure is chosen
to reproduce the field-theory propagator.

It is now straightforward to compute the expectation value of the circular
loop using classic results in random matrix theory \cite{Brezin:1977sv}.
The eigenvalues of the Gaussian random matrix $M$ have
a continuous distribution with finite support in the large-$N$
limit. The distribution of eigenvalues obeys the semi-circle law:
\eq{
\vev{\frac{1}{N}\,\tr f(M)}_M=\frac{2}{\pi}\int_{-\sqrt{\lambda}}^{\sqrt{\lambda}}
dm\,\sqrt{\lambda-m^2}\,f(m).
}
Substituting $f(m)=\e^m$, we find:
\eq{\label{exactcir}
\vev{W(\cir)}=\frac{2}{\sqrt{\lambda}}\,I_1\br{\sqrt{\lambda}},
}
where $I_1$ is modified Bessel function.

We can compare this result with the prediction of AdS/CFT
correspondence by taking the large-$\lambda$ limit: \eq{
\vev{W(\cir)}=\sqrt{\frac{2}{\pi}}\,\lambda^{-3/4}\e^{\sqrt{\lambda}}
~~~~~(\lambda\rightarrow\infty).
} The prediction of the string theory, eq.~\rf{adswl1}, has exactly
the same form.  Recalling that the area of minimal surface associated
with the circle is equal to $-2\pi$, we find the complete agreement with
string theory prediction! The exact expression \rf{exactcir}
smoothly interpolate between perturbative series in $\lambda$ and the
strong coupling regime, where the natural expansion parameter is $1/\sqrt{\lambda}$.
This latter expansion is to be identified with $\alpha'$ expansion
of the world-sheet sigma model.

The summation of rainbow diagrams for the circular Wilson loop can be
extended to all orders of $1/N^2$ expansion. In agreement with
expectations from string theory, each order contains the same
exponential factor multiplied by an overall power of $\lambda^{1/4}$
at strong coupling \cite{Drukker:2000rr}: \eq{
\vev{W(\cir)}=\sqrt{\frac{2}{\pi}}\,
\sum_{g=0}^{\infty}\frac{1}{N^{2g}}\,\frac{1}{96^g
g!}\,
\lambda^{(6g-3)/4}\e^{\sqrt{\lambda}}~~~~~(\lambda\rightarrow\infty).
} As was explained by Drukker and Gross \cite{Drukker:2000rr}, the
power of $\lambda^{1/4}$ at $g$-th order of $1/N^2$ expansion
correctly counts the number of zero modes at $g$-th order of string
perturbation theory.

\subsection{OPE coefficients for chiral primary operators}

At weak coupling, the OPE coefficient of the circular Wilson loop
for dimension-$k$ CPO \rf{cpo}
is proportional to $\lambda^{k/2}$:
\begin{equation}\label{wc}
\frac{\langle W(\cir)\,{\cal O}^I_k\rangle }{\langle W(\cir)\rangle 
}=2^{-k/2}\,\frac{\sqrt{k}}{k!}\,\lambda^{k/2}\,\frac{R^k}{L^{2k}}
Y^I(\theta) +\ldots~~~(\lambda\rightarrow 0).
\end{equation}
Comparing this with the AdS/CFT prediction \rf{sc} we see that OPE
coefficients are non-trivial functions of the 't~Hooft coupling.

\begin{figure}[h]
\hspace*{4cm}
\epsfxsize=8cm
\epsfbox{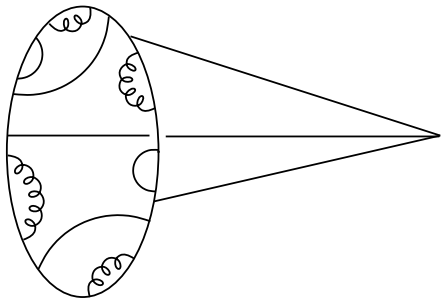}
\caption[x]{\small A typical diagram that contributes to the correlator
of the circular Wilson loop with CPO.}
\label{diagr}
\end{figure}

Again, appealing to supersymmetry and conformal invariance,
we argue that  correlators of the circular Wilson loop
with chiral operators
are saturated by free fields.  Therefore, calculation of these
correlators amounts in resummation of all planar rainbow diagrams of
the kind shown in fig. \ref{diagr}. This is a rather lengthy exercise
for arbitrary $k$ which involves the use of loop equations
\cite{Migdal:1983gj,Makeenko:tb,Akemann:2001st} in the matrix model
\rf{rmm}. The details may be found in the original reference
\cite{Semenoff:2001xp}. Here, we only quote the result:
\begin{equation}\label{ex}
\frac{\langle W(\cir)\,{\cal O}^I_k\rangle}{\langle W(\cir)\rangle
}=2^{k/2-1}\sqrt{k\lambda}\,
\,\frac{I_k\left(\sqrt{\lambda}\right)}{I_1\left(\sqrt{\lambda}\right)}\,
\frac{R^k}{L^{2k}}\,Y^I(\theta).
\end{equation}
We expect that this expression is exact in the large N limit.  Its
perturbative series expansion starts with (\ref{wc}).

At strong coupling we expect to reproduce the AdS/CFT prediction
\rf{sc}, and this is indeed the case, because all modified Bessel
functions have the same asymptotics at infinity.
This provides an infinite series of correlation functions, for which
resummed perturbative series allow to trace an interpolation between
weak coupling regime and the strong-coupling prediction of string
theory in Anti-de-Sitter space.
 
The non-perturbative expression \rf{ex} resolves the puzzle
mentioned in sec.~\ref{wlc}, where we argued that OPE coefficients
must be small for operators of large dimension and noticed that this
is not the case if we use the supergravity prediction for OPE
coefficients. 
However, expanding the Bessel function $I_k\left(\sqrt{\lambda}\right)$
in $\lambda$ one can check 
that the smallness parameter of perturbation theory for
$\langle W(C)\,{\cal O}^I_k\rangle$ is not $\lambda$, but $\lambda/k$,
so large-$k$ limit is always perturbative. If we take the limit
$k\rightarrow\infty$ before $\lambda\rightarrow\infty$, 
we can keep only the first term of perturbation series,
which is indeed suppressed by $1/k!$.

\newsection{Remarks}

One of the many achievements made possible by the discovery of the
AdS/CFT correspondence is a systematic way to do computations in the 
interacting field theory at strong coupling.  These computation
are done by methods that are quite unusual 
and sometimes counterintuitive 
from the field theory perspective. It is therefore very important that
non-trivial predictions of string theory and supergravity can be reproduced
by more or less ordinary techniques of planar perturbation theory.
Of course, this is possible only in special cases and depends on symmetries
of \N SYM theory, but the very fact that it is possible is quite surprising.
It is also important that exact field-theory calculations can be done
for Wilson loops which probe string theory directly and therefore allow to 
test the AdS/CFT correspondence in its strongest form.

The current status of this subject leaves many questions unanswered.
Some of the immediate questions are

\noindent$\bullet$The straight line Wilson loop appears to have unit
expectation value.  This is a prediction of the supergravity
computation for the strong coupling limit and it also seems to be so
for perturbative computations to a reasonably high order in the
Feynman gauge.  In other gauges which are related by conformal
transformation with the Feynman gauge, the leading perturbative
corrections need not vanish but can reproduce the perturbative limit
of the circle Wilson loop.  It is clear that a deeper understanding of
the gauge dependence of this object is needed.  It would be
interesting to extend the arguments for non-renormalization of
correlators of local BPS operators to the case of the Wilson 
line, which is a non-local operator.

\noindent$\bullet$ There should be a more rigorous proof that
radiative corrections to the results in this paper actually cancel.
One approach which was suggested in \cite{Drukker:2000rr} is to show
that the result for the circle comes from a conformal
anomaly. Establishing this at a rigorous level would be an important
step in the right direction.

\noindent$\bullet$ It should be possible to study other kinds of
Wilson loops 
\cite{Erickson:2000qv,
Erickson:1999uc,Plefka:2001bu,Arutyunov:2001hs}.  

\noindent$\bullet$  Most desirable would be to obtain some results for
non-conformally invariant gauge theories.  At this point this appears
to be very difficult as most of the analytic computations that have
been done so far depend heavily on conformal invariance.


\end{document}